\documentclass[prd, twocolumn, nofootinbib, floatfix]{revtex4-1}

\usepackage{amsmath}
\usepackage{graphicx}
\usepackage{dcolumn}
\usepackage{bm}
\usepackage{epsfig}
\usepackage{amssymb,latexsym,mathrsfs}
\usepackage{graphicx}
\usepackage{color}
\usepackage{hyperref}

\hypersetup{
    colorlinks=true,
    linkcolor=red,
    citecolor=blue,
} 

\newcommand{\be}{\begin{equation}}
\newcommand{\ee}{\end{equation}}
\newcommand{\bs}{\begin{split}} 
\newcommand{\bea}{\begin{eqnarray}}
\newcommand{\eea}{\end{eqnarray}}

\newcommand{\om}{\Omega_m}

\newcommand{\dl}{\delta}

\begin{document}

\title{Doubling Strong Lensing as a Cosmological Probe} 
\author{Eric V.\ Linder} 
\affiliation{Berkeley Center for Cosmological Physics \& Berkeley Lab, 
University of California, Berkeley, CA 94720, USA} 

\begin{abstract}
Strong gravitational lensing provides a geometric probe of cosmology in a 
unique manner through distance ratios involving the source and lens. 
This is well known for the time delay distance derived from measured delays 
between lightcurves of the images of variable sources such as quasars. 
Recently, double source plane lens systems involving two constant sources 
lensed by the same foreground lens have been proposed as another probe, 
involving a different ratio of distances measured from the image positions 
and fairly insensitive to the lens modeling. Here we demonstrate 
that these two different sets of strong lensing distance ratios have 
strong complementarity in cosmological leverage. 
Unlike other probes, the double source distance ratio is actually more 
sensitive to the dark energy equation of state parameters $w_0$ and $w_a$ 
than to the matter density $\om$, for low redshift lenses. Adding double 
source distance ratio measurements can improve the dark energy figure of merit 
by 40\% for a sample of fewer than 100 low redshift systems, or even better 
for the optimal redshift distribution we derive. 
\end{abstract} 

\date{\today} 

\maketitle

%%%%%%%%%%%%%%%%%%%%%%%%%%%%%%%%%%%%% 
\section{Introduction} 

The concept of strong gravitational lensing, the significant impact of 
gravity on the propagation of light, dates back to Michell in 1783 
\cite{michell}. Michell used this to predict black holes (then called 
dark stars) and gravitational redshift. In modern cosmology, strong 
lensing can be used a geometric probe of spacetime and the expansion of 
the universe. The distance ratios, similar to focal lengths, entering 
in strong lensing effects such as multiple image separations and time 
delays provide complementary information to the expansion history mapping 
from single distances such as the luminosity distance of supernovae or 
angular distance of baryon acoustic oscillations. 

Three main aspects of strong lensing have been used as cosmological probes: 
the statistical abundance of strongly lensed images, whether arcs or multiple 
images, the angular separation of multiple images in individual systems, 
and the time delay 
between multiple images of a variable source. Each involves different 
ratios of the distances between the observer and lens, observer and source, 
and lens and source. The sensitivity of these to cosmology, especially 
the dark energy equation of state, and their unique properties relative to 
single distances, was discussed by \cite{lin04}. In particular, the last 
two observables can exhibit positive correlation between the dark energy 
equation of state value today $w_0$ and a measure of its time variation $w_a$, 
unlike the classic cosmology probes. This offers the potential for 
complementarity with standard, single distance measurements and hence 
greater leverage on cosmology estimation. 

Each strong lensing technique has its specific dependence on other 
ingredients besides cosmographic distances, e.g.\ selection effects and 
the growth of cosmic structure in the case of abundances, 
or the mass profile of the lens and line of sight structure in the last 
two methods. Time delay cosmography has made the greatest advances in the 
last several years, with improved lens modeling 
\cite{oguri07,suyu09,suyu10,suyu12,suyu13,collett,greene}, clear 
understanding of the cosmological leverage \cite{lin11,treu13}, high accuracy 
time delay estimation \cite{hkl,tdc0,tdc1,hoj,amir,tewes,tak}, and actual 
application to cosmological constraints on geometry and dark energy 
\cite{suyu12,suyu13,holicow}. For an up to date review, see \cite{tm16}. 

In this article we return to investigation of the second strong lensing 
technique, using image separations. This is also a geometric probe of 
cosmology, and recent developments have increased its potential. The 
sensitivity to the lensing mass distribution poses an obstacle to its use 
for precision cosmology, but \cite{coll1,coll2} proposed using a ratio of 
ratios technique for canceling much of this dependence. Double source 
plane lensing, where two independent sources are each lensed into multiple 
images by the same foreground mass, involves a ratio of distance ratios 
where the lens model nearly cancels out. Moreover, examples have recently 
been observed and the number of such systems is likely to increase with 
wide field surveys underway such as the Dark Energy Survey, HyperSuprime Cam, 
and KiloDegree Survey. 

In Sec.~\ref{sec:method} we briefly summarize the concept of double 
source plane lensing (DSPL) and investigate the cosmological sensitivity of 
its central quantity, the ratio of distance ratios. We demonstrate the 
significant complementarity of DSPL measurements with time delay lens 
observations in Sec.~\ref{sec:cos}, in terms of constraints on dark energy 
and cosmology. Section~\ref{sec:zdist} examines the impact of the redshift 
distribution of the lens and source sample, relevant to future surveys such 
as Euclid, LSST, and WFIRST, and Sec.~\ref{sec:concl} presents the conclusions.

%%%%%%%%%%%%%%%%%%%%%%%%%%%%%%%%%%%%%%%%%  
\section{Double Source Plane Lensing Distances} \label{sec:method} 

\subsection{Introduction to DSPL} \label{sec:dsplintro} 

A double source plane lensing system occurs when two background sources 
are sufficiently aligned behind a common foreground lens that both are 
split into multiple images. The angular separation between images depends 
on the ratio of the lens-source distance $D_{ls}$ and observer-source distance 
$D_s$, and the mass of the lens (feeding into its Einstein radius). 
In the ratio of angular separations for images from source 1 
and images from source 2, the lens mass will cancel out (in the ideal case), 
leaving a purely geometric distance measure. 

There are subtleties to this picture, treated in \cite{coll1,coll2,gavazzi}. For 
example the lens model cancellation is exact only for a singular isothermal 
sphere distribution -- but in the general case the lens model uncertainty is 
significantly suppressed. A form of the mass sheet degeneracy common to 
lensing systems, where mass apart from the lens can be degenerate with the 
observables, exists for DSPL \cite{schn14} and requires dynamical information 
(e.g.\ lens velocity dispersion) to break. (Also see \cite{schn16} regarding 
weak lensing shear ratios for multiplane sources and lenses.) 
Lensing of the further source by 
the nearer one must also be taken into account. Finally, double source lensing 
is rarer than single source lensing so there are fewer systems to use in 
cosmological constraints, but on the other hand there is no requirement for 
the sources to be time variable and monitored for extended periods of time 
as for time delay lensing. 

Just as has been done for time delay cosmography, 
all these sources of systematic uncertainty will need to be addressed through 
a program of observations, data simulation challenges, and theoretical work. 
To motivate that such effort is worthwhile, here we investigate the 
cosmological constraint impact if DSPL becomes a new, mature probe. 
This is particularly relevant with the approach of next generation wide 
field surveys such as Euclid, LSST, and WFIRST that should find abundant 
samples of DSPL. For example, \cite{gavazzi} estimates that the Euclid 
satellite will find of order $\sim$2000 DSPL and the ground based Large 
Synoptic Survey Telescope (LSST) will find a similar number. Using the 
technique of \cite{coll15}, \cite{collpriv} estimates that using only the 
best seeing exposures to get the highest quality will deliver $\sim160$ DSPL 
from Euclid and $\sim120$ from LSST. The WFIRST 
satellite will find fewer but has the advantage of excellent spatial 
resolution for more precise image separation measurements. 

Initial work on cosmological parameter estimation appears in 
\cite{coll1,coll2}, concentrating on a flat universe with constant dark 
energy equation of state $w$ and a restricted sample of lenses. We expand 
on this by focusing on the more general time varying dark energy equation 
of state $w(z)=w_0+w_a\,z/(1+z)$ in common use, and exploring the 
influence of distributions in the redshifts of lens, source 1, and source 2, 
and briefly examining the case of nonflat universes. 
Most importantly, we identify 
strong complementarity between the two strong lensing probes of DSPL and 
time delay lensing.

%%%%%%%%%%%%%%%%%% 
\subsection{Cosmological sensitivity} \label{sec:sens} 

The central quantity for DSPL is the ratio of distance ratios, 
\be 
\beta=\frac{D_{ls}(z,z_1)}{D_s(z_1)}\frac{D_s(z_2)}{D_{ls}(z,z_2)}\ , 
\ee 
where the lens is at redshift $z$, the nearer source is at $z_1$, and the 
further source at $z_2$. We begin by examining its sensitivity to 
cosmological parameters. 

Most obviously, it is dimensionless and so independent of the Hubble 
constant $H_0$. This alone indicates some complementarity with the dimensional 
time delay distance, which is the ratio 
\be 
D_{\Delta t}=\frac{D_l(z)\,D_s(z_1)}{D_{ls}(z,z_1)}\,(1+z)\ , 
\ee 
and so inversely proportional to $H_0$. To probe the sensitivity of $\beta$ 
to the time varying dark energy equation of state parameters, we use the 
Fisher information formalism. Note that $\beta=\beta(z,z_1,z_2)$ so for 
visual ease of presentation we exhibit the results for a series of lens 
redshifts $z$, and set $z_1/z=2$, $z_2/z_1=1.5$. This is motivated by the 
lensing kernel peaking when the lens is roughly halfway between the source 
and observer, but we analyze various other cases both in this section and in 
Sec.~\ref{sec:zdist}. 

\begin{figure}[htbp!]
\includegraphics[width=\columnwidth]{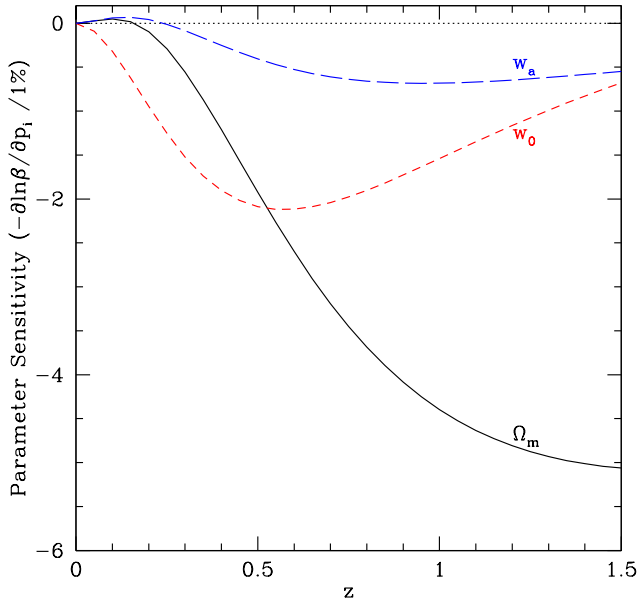} 
\caption{
The sensitivity of measurements of the double source lensing distance ratio 
$\beta$ for constraining cosmological parameters is plotted as a function 
of the lens redshift $z$. The magnitude of the sensitivity is for a 1\% 
measurement of $\beta$, but the more interesting aspects come from the 
shape of the curves: the null of the $\om$ curve at $z\approx0.15$ and the 
opposite signs for $w_0$ and $w_a$ sensitivities for $z\lesssim0.23$, 
indicating a distinct behavior from single distance probes. 
} 
\label{fig:sens} 
\end{figure}

Figure~\ref{fig:sens} illustrates the cosmological sensitivity of $\beta$ 
as a function of $z$ through the Fisher derivatives 
$\partial\beta/\partial p_i$, for the parameters $p_i$ of the dimensionless 
matter density $\om$, the dark energy equation of state today $w_0$, and 
the dark energy equation of state time variation $w_a$. The fiducial 
cosmology is a flat $\Lambda$CDM universe with $\om=0.3$. (We assume spatial 
flatness throughout the article, except for part of Sec.~\ref{sec:cos} that 
examines the free curvature case.) The larger the 
absolute value of the sensitivity, and the greater the distinction between 
the shapes of the derivative curves, the more information exists and the 
tighter are the cosmological constraints.

%%%%%%%%%%%%%%%%%%% 

We identify several interesting properties. Classic cosmology probes are 
more sensitive to the matter density $\om$ than the dark energy equation of 
state (for example, low redshift supernova distances probe the deceleration 
parameter $q_0=[1+3w(1-\om)]/2$, so the sensitivity derivative with respect 
to $\om$ is $\sim$1.4 times larger than that for $w$, and this only gets 
larger at higher redshift). DSPL, however, is more sensitive to both $w_0$ 
and $w_a$ than $\om$ at low redshift, and indeed has a null in the dependence 
on $\om$ at $z\approx0.15$. This means that degeneracy with the matter 
density is strongly suppressed there. Furthermore, $w_0$ and $w_a$ have 
opposite signs for $z<0.23$, meaning they have positive correlation, unlike 
single distance and growth probes. The sensitivity to dark energy parameters 
is relatively strong out to $z\approx0.6$, before the matter density 
dominates for higher redshift systems. This means that the systems with the 
greatest leverage are at observationally benign low redshifts, easing 
followup requirements such as redshift or velocity dispersion measurements. 
Thus, DSPL appears at first glance to be quite promising. 

\begin{figure}[htbp!]
\includegraphics[width=\columnwidth]{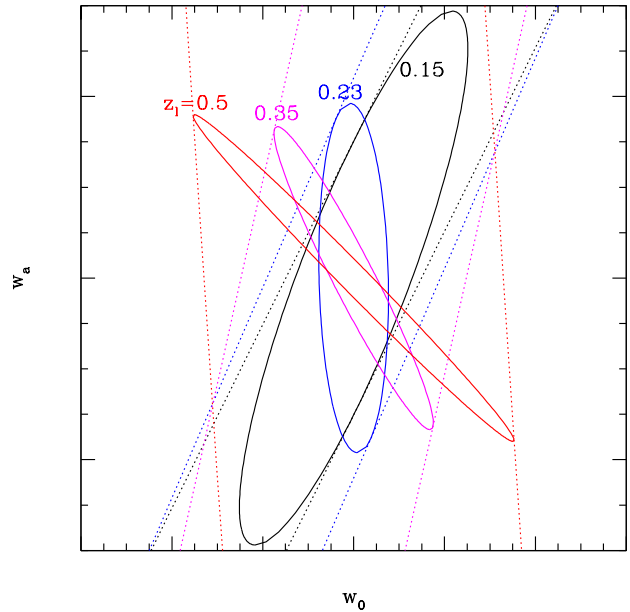} 
\caption{
The leverage of measurements of the double source lensing distance ratio 
$\beta$ for constraining the dark energy equation of state value today $w_0$ 
and time variation $w_a$ is plotted for different lens redshifts $z$. This 
shows the evolution of covariance direction. Solid ellipses fix $\om=0.3$ 
while dotted ellipses marginalize over $\om$. 
} 
\label{fig:flower} 
\end{figure}

Taking into account the covariances between the cosmological parameters, 
we can use Fisher information analysis to estimate constraints on the 
cosmology for a series of measurements. To emphasize the unique aspects 
of positive correlation between $w_0$ and $w_a$, and reduced sensitivity 
to the matter density degeneracy, we show an illustrative calculation (omitting 
the axis values due to the idealized precision) in Fig.~\ref{fig:flower}.

%%%%%%%%%%%%%%%%%%% 

The solid ellipses highlight the evolution of the degeneracy direction 
between the dark energy equation of state parameters by fixing $\om=0.3$. 
We see that indeed the case with the lens redshift at $z=0.23$, where the 
$w_a$ curve in Fig.~\ref{fig:sens} crosses zero, gives a vertical 
parameter estimation contour (we actually use lenses at $z$, $z\pm0.05$ 
to obtain finite contours). Lower redshift lenses, where the $w_0$ and $w_a$ 
sensitivities are positively correlated, give ellipses leaning to the right, 
the opposite of standard single distance and growth probes, while higher 
redshift lenses give negative correlation. The dotted lines show the 
elongation of the contours when marginalizing over $\om$, as should be done. 
The expansion of the contours due to the matter density covariance is most 
severe at higher redshift, and changes the contour direction and size 
less at lower redshift. Interestingly, the marginalization over $\om$ 
actually tilts the ellipses to the right, so positive correlation in the 
$w_0$-$w_a$ plane occurs for lenses at $z\lesssim0.4$, not just $z<0.23$.

%%%%%%%%%%%%%%%% 
\subsection{Analytic redshift dependence} \label{sec:zdep} 

DSPL seems an interesting possibility as a new complementary cosmological 
probe therefore. Before proceeding further, let us return to the point 
about the source redshifts. Consider moving the further source closer to 
the nearer one, say $z_2=z_1+0.1$. This dramatically reduces the cosmological 
sensitivity, by a factor $\sim$10 according to numerical computation. 
We can derive this analytically. 
Let $z_2=z_1+\dl$. Then 
\bea 
D_s(z_2)&=&(1+z_2)^{-1}\int_0^{z_2} dz/H(z)\notag\\ 
&=&(1+z_1+\dl)^{-1}\left[(1+z_1)D_s(z_1)+\dl/H(z_1)\right]\notag\\ 
&\approx&D_s(z_1)\,\left[1+\frac{\dl}{1+z_1}\,\left(\frac{1}{H_1D_1}-1\right)\right]\ . 
\eea 
Calculating $D_{ls}(z,z_2)$ similarly, this gives 
\be 
\beta\approx 1+\frac{\dl}{1+z_1}\left[\frac{1}{H_1D_1}-\frac{1}{H_1D_{ls}(z,z_1)}\right] \ . 
\ee 
This makes sense since as $z_2/z_1\to1$, then $\beta\to1$ and a constant 
has no cosmological sensitivity. Thus the sensitivity is suppressed by 
the small factor $\dl/(1+z_1)$. 

What if we take $z_2/z_1\gg1$? Very high redshift sources are harder to 
observe since their images will generally be fainter; also, the followup 
resources needed to measure their redshift will be more expensive. Thus, 
we take the somewhat conservative case of using $z_2/z_1=1.5$ as a baseline, 
though we revisit this in Sec.~\ref{sec:zdist}. 

If we were to move the closer source nearer to the lens, we reduce the 
value of $\beta$. Instead of it being near unity, it can become much 
smaller. Again, this reflects that the peak of the lensing kernel is 
where the lens is roughly midway between the source and observer. For 
$z_1=z+\dl$, 
\be 
\beta\approx \frac{\dl}{1+z}\,\frac{1}{H(z)D(z)}\,\frac{D_s(z_2)}{D_{ls}(z,z_2)} 
\ . 
\ee 
We see this is suppressed by $\dl/(1+z)$. However, an interesting point 
is that we also have sensitivity not just to distances but to the Hubble 
parameter $H(z)$ directly. Which one wins out in its influence on the 
cosmological parameter estimation depends on the measurement assumptions. 
If one says the measurements are at constant relative precision, e.g.\ 
1\% measurements, then we find that the extra information from the direct 
$H(z)$ dependence is more significant. Note though that since $\beta$ may 
be smaller by a factor 10, say, this means that the measurement must have 
$\sigma_\beta\sim0.001$ rather than $\sigma_\beta\sim0.01$, for a 1\% 
measurement of $\beta$. This would be highly challenging, and therefore 
we conservatively take $z_1/z=2$, although again we revisit this in 
Sec.~\ref{sec:zdist}.

%%%%%%%%%%%%%%%%%%%%%%%%%%%%%%%%%%%%%%%%%  
\section{Cosmological Constraints} \label{sec:cos} 

While DSPL has interesting cosmological sensitivity properties, it does 
not have much raw sensitivity magnitude. Rather it is the complementarity 
with time delay lensing and a high redshift probe such as the cosmic 
microwave background (CMB) that is of value. We consider time delay lensing 
as in \cite{lin11} -- 1\% precision on $D_{\Delta t}$ in six redshift bins from 
$z=0.1$-0.6 -- and CMB measurements of the distance to last scattering 
and the physical matter density $\om h^2$ (where $h=H_0/100$ km/s/Mpc) of 
the quality of Planck satellite measurements. For DSPL we adopt as the 
baseline 1\% measurements, basically of the image separations, of 96 systems 
in the range $z=0.1$-0.6. Recall that this range was identified as most 
theoretically promising, as well as observationally tractable, in 
Sec.~\ref{sec:sens}. 

\begin{figure}[htbp!]
\includegraphics[width=\columnwidth]{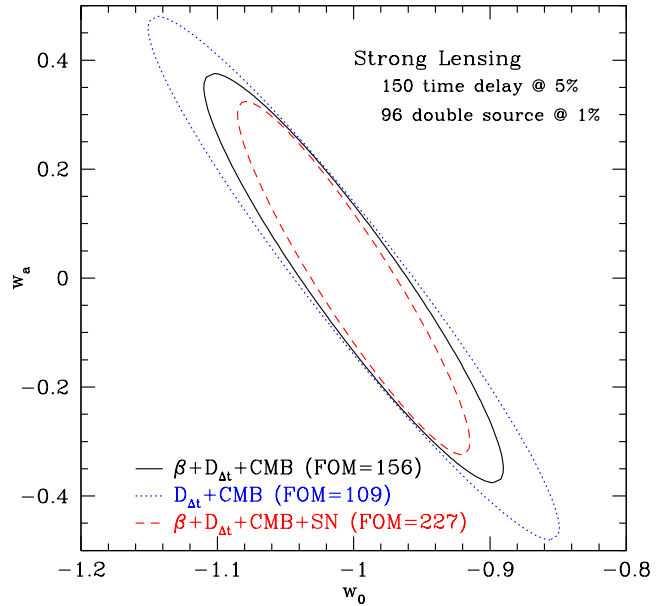} 
\caption{
Cosmological parameter estimation uncertainty is plotted in the dark energy 
equation of state plane for the case of strong lensing time delays (dotted 
curve), time delays plus double source plane strong lensing (solid), and the 
two strong lensing probes plus supernova distances (dashed). DSPL brings 
significant complementarity. 
} 
\label{fig:ellcos} 
\end{figure}

The addition of this new probe improves the dark energy figure of merit 
FOM=$\sqrt{\det F_{w_0w_a}}$, a measure of the uncertainty in dark energy 
equation of state estimation, by 43\%. Figure~\ref{fig:ellcos} shows the 
confidence contours in the $w_0$-$w_a$ plane, marginalized over the other 
cosmological parameters. The matter density $\om$ is determined to 0.0047, 
$w_0$ to 0.072, $w_a$ to 0.25, and the reduced Hubble constant $h$ to 0.0047.

%%%%%%%%%%%%%%%%%%% 

Thus both strong lensing probes work well together. Furthermore, the 
combination remains complementary with a single distance probe like 
supernovae. The dashed curve shows the addition of a midterm supernova 
sample of the rough quality, including systematics, expected by the end 
of the Dark Energy Survey, approximately equivalent to 150 supernovae at 
$z<0.1$, 900 at $z=0.1-1$, and 42 at $z=1-1.7$ (as used in \cite{lin11}), 
with a systematic of $0.02(1+z)/2.7$ mag. 

Degrading the DSPL precision to 2\% reduces the FOM by 22\%. 
We can also examine the impact of the redshift range used for the lenses. 
Cutting out the low ($z=0.1$) or high ($z=0.6$) part of the sample reduces 
the FOM by 14\%, while extending it to $z=0.7$ improves it by 13\%. As 
expected, estimation of $h$ suffers most when removing the low redshift 
systems, while predominantly the dark energy parameter constraints loosen 
when removing the high redshift systems. 
Thus the range of lens system redshifts between 
$z=0.1-0.6$ seems a happy medium, especially as higher redshift systems 
become more expensive for followup to obtain spectroscopic redshifts and 
lens velocity dispersions. 
(Note that \cite{linopt} describes code for optimizing the science 
return under constrained followup resources by a merit vs cost weighting.) 

\begin{figure}[htbp!]
\includegraphics[width=\columnwidth]{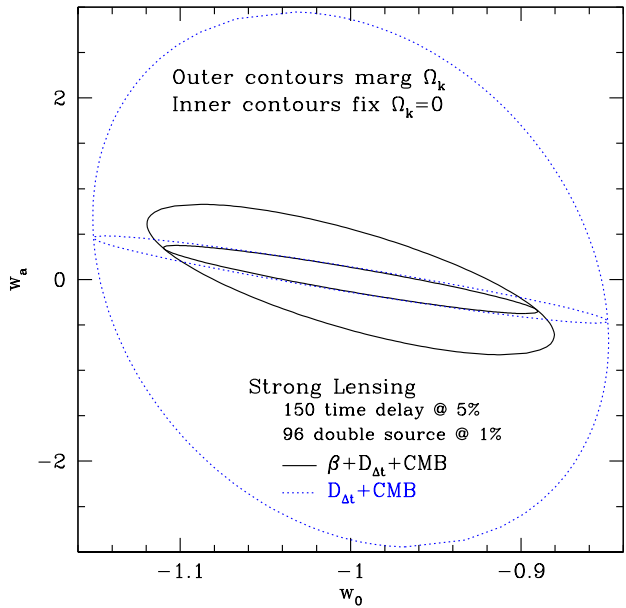} 
\caption{
Cosmological parameter estimation uncertainty, allowing for (outer contours) 
and fixing (inner contours) spatial curvature, is plotted in the dark energy 
equation of state plane. The curvature induced degeneracy in the case of 
strong lensing time delays (dotted curves) is substantially tamed by the 
combination of double source plane lensing with the lensing time 
delay measurements (solid curves). 
} 
\label{fig:curv} 
\end{figure}

Recalling that \cite{lin11} found that time delay lensing greatly immunized 
supernova distances against the degeneracy due to spatial curvature, we 
consider the effect of doubling strong lensing. Figure~\ref{fig:curv} shows 
that combining DSPL with time delay lensing similarly reduces the constraint 
loss due to spatial curvature density $\Omega_k$. While allowing $\Omega_k$ 
to float blows up the $w_0$--$w_a$ contour area by a factor 20, relative to 
the flat case, for time delay lensing plus CMB, this factor is only 4.5 for the 
DSPL plus time delay lensing plus CMB combination. Moreover, in the curvature 
free case, including DSPL tightens the constraint on the curvature by a 
factor 4.6, to $\sigma(\Omega_k)=0.0072$, and tightens the $w_a$ determination 
by a factor 3.6.

%%%%%%%%%%%%%%%%%%% 

%%%%%%%%%%%%%%%%%%%%%%%%%%%%%%%%%%%%%%%%%  
\section{Redshift Sensitivity} \label{sec:zdist} 

Although we gave good rationales in Sec.~\ref{sec:method} for why the 
choices $z_1/z=2$ and $z_2/z_1=1.5$ were reasonable, from both 
theoretical sensitivity and observational followup points of view, 
let us revisit the question of optimal redshift distributions. 
We vary these two redshift ratios and study the 
impact on dark energy figure of merit. 

\begin{figure}[htbp!]
\includegraphics[width=\columnwidth]{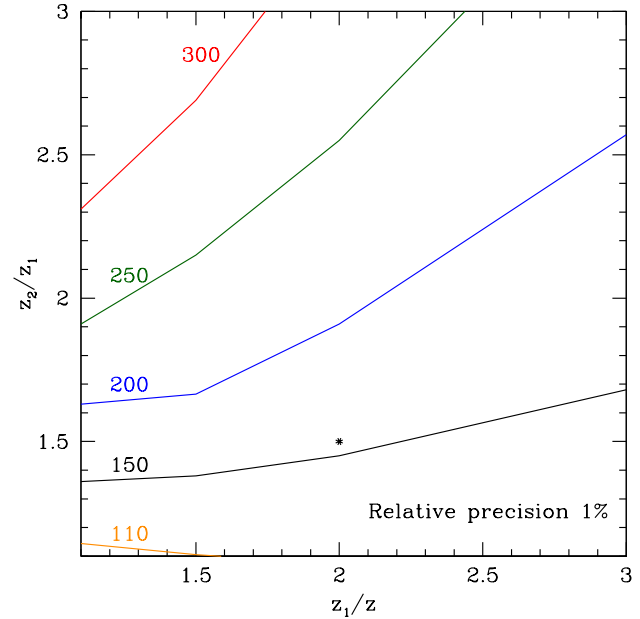} 
\caption{Isocontours of dark energy figure of merit are plotted for 
a range of source-source and source-lens redshift ratios, assuming constant 
relative precision of 1\% on $\beta$. The baseline case used in 
Sec.~\ref{sec:cos}, justified in Sec.~\ref{sec:method}, with $z_1/z=2$ and 
$z_2/z_1=1.5$, is indicated by the star. 
} 
\label{fig:varyzrel} 
\end{figure}

Figure~\ref{fig:varyzrel} shows the results for the case of constant 
relative precision, i.e.\ 1\% per DSPL system. Recall from 
Sec.~\ref{sec:method} that we showed analytically that as $z_1/z$ gets 
close to unity, $\beta$ involves the Hubble parameter $H(z)$ itself. 
This would be expected to bring in extra cosmological information and 
indeed that is exactly what we find: the figure of merit improves as 
we move to the left in the figure. As we increase $z_2/z_1$ and move up 
in the figure, we have an increased lever arm in distance, allowing for 
greater complementarity in the measurements, and again the FOM increases. 
For one source close to the lens and the other much further away, the 
FOM from the combination of DSPL, time delay lensing, and CMB can reach 
373.

%%%%%%%%%%%%%%%%%%% 

However, this obscures some difficulties. Pushing the ratios in 
these directions lowers $\beta$, and it is only the assumption of constant 
relative precision that enables such gains. Indeed in the extreme case 
leading to a FOM of 373, we have $\beta=0.13$ (rather independent of lens 
redshift) and so are positing an extremely small measurement uncertainty 
$\sigma_\beta=0.0013$. Moreover, the further source will be at a high redshift 
increasingly difficult for precision followup. 

\begin{figure}[htbp!]
\includegraphics[width=\columnwidth]{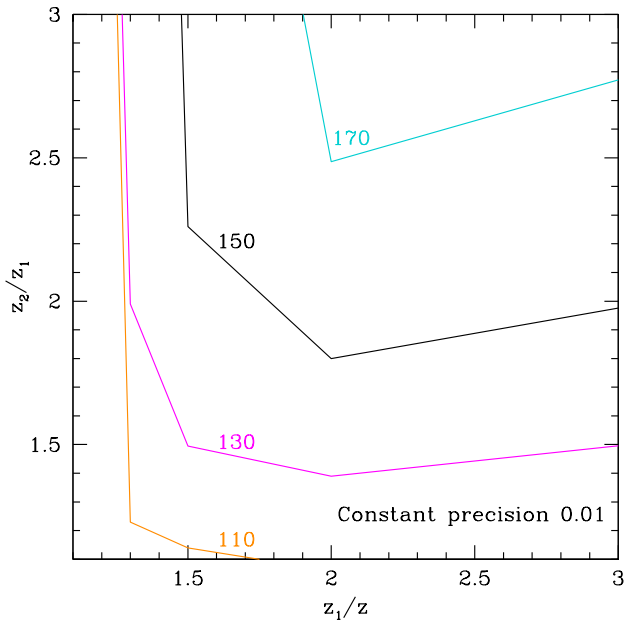} 
\caption{Isocontours of dark energy figure of merit are plotted for 
a range of source-source and source-lens redshift ratios, as in 
Fig.~\ref{fig:varyzrel}, but now assuming a more realistic constant 
absolute precision of 0.01 on $\beta$ (corresponding to $\sim$1\% relative 
precision for well spaced sources and lens). 
} 
\label{fig:varyzcon} 
\end{figure}

Instead let us consider constant absolute precision $\sigma_\beta=0.01$. 
This will penalize redshift distributions that give low $\beta$, since then 
the signal to noise is lower. Figure~\ref{fig:varyzcon} bears this out 
exactly. Indeed we see that the optimum source 1 to lens redshift ratio 
$z_1/z=2$, in order to get the highest FOM for fixed source 2 to source 1 
redshift ratio $z_2/z_1$. This justifies our previous adoption of $z_1/z=2$. 
As one pushes the farther source to higher redshift, $\beta$ slowly declines 
from unity with increasing $z_2/z_1$, but the longer cosmological lever arm 
wins out and the FOM increases. The price of this, however, is more difficult 
observations: the images of the farther source will be fainter and of lower 
signal to noise and hence harder to measure well, plus followup time for, 
e.g., the source redshift will be more expensive. We therefore have used 
the conservative assumption that $z_2/z_1=1.5$. Further work, using specific 
exposure time calculations for a given survey instrument, may eventually 
indicate that this source-source redshift ratio can be increased; this would 
bring a further gain in FOM.

%%%%%%%%%%%%%%%%%%% 

%%%%%%%%%%%%%%%%%%%%%%%%%%%%%%%%%%%%%%%%%  
\section{Conclusions} \label{sec:concl} 

Cosmography is an important element of the quest to understand the 
nature of cosmic acceleration, especially to the extent that it is free 
of uncertainties in the growth of structure (e.g.\ the mapping between 
galaxies or clusters and the underlying dark matter field). A new 
geometrical probe would be a valued addition to the cosmic toolbox. 
While it is very early days yet, double source plane lensing (DSPL) 
possesses several intriguing properties that motivate further development 
of theoretical, simulation, and observational studies. 

Uniquely, DSPL for low redshift systems is more sensitive to the dark 
energy equation of state than to the matter density, and indeed there is 
a nulling of the matter density degeneracy. Furthermore, it is one of the 
rare probes, involving distance ratios like time delay lensing, that has 
a positive correlation between dark energy equation of state parameters 
$w_0$ and $w_a$, thus offering special complementarity with single distance 
measures like supernovae and baryon acoustic oscillations. Finally, we have 
demonstrated that double source plane lensing 
has complementarity with time delay lensing, doubling 
strong gravitational lensing as a cosmological probe. 
This holds in both flat and, especially, spatial curvature free cosmologies. 

We have identified the optimum redshift distributions of lens and sources 
under various measurement assumptions, both analytically and numerically, 
and quantified the dark energy figures of merit. Improvement of the figure 
of merit by 40\% with 
the addition of DSPL is found under reasonable assumptions. Issues of 
systematics, and observational and followup practicalities, 
certainly remain to be addressed, but the calculations here show the worth 
of undertaking such efforts. 

Double source plane lensing has become an observational reality, with 
two galaxy scale systems currently known, dozens more likely to be found by 
the current generation of wide field surveys, and hundreds to thousands 
expected in the next generation by Euclid, LSST, and WFIRST. Time delay 
lensing is similarly poised for a cornucopia of data. Doubling 
strong lensing may prove to be a fruitful path forward in understanding 
the geometry and accelerated expansion of the universe.

%%%%%%%%%%%%%%%%%%%%%%% 
\acknowledgments 

I thank the Institute of Cosmology and Gravitation, University of Portsmouth 
for hospitality during the ICG-KASI workshop that 
inspired this work. This work is supported in part by the U.S.\ Department 
of Energy, Office of Science, Office of High Energy Physics, under Award 
DE-SC-0007867 and contract no.\ DE-AC02-05CH11231.

%%%%%%%%%%%%%%%%%%%%%%%%%%%%%%%% 

\end{document}